# Physics Design of the ETA-II/Snowtron Double Pulse Target Experiment*

Yu-Jiuan Chen, Darwin D.-M. Ho, James F. Mccarrick, Arthur C. Paul, Stephen Sampayan,
Li-Fang Wang, John T. Weir, LLNL, Livermore, CA 94550, USA

*Abstract*

We have modified the single pulse target experimental facility[1] on the Experimental Test Accelerator II (ETA-II) to perform the double pulse target experiments to validate the DARHT-II[2, 3] multi-pulse target concept. The 1.15 MeV, 2 kA Snowtron injector will provide the first electron pulse. The 6 MeV, 2 kA ETA-II beam will be used as the probe beam. Our modeling indicates that the ETA-II/Snowtron experiment is a reasonable scaling experiment.

## 1 INTRODUCTION

The DARHT-II facility will provide four 2.1 mm spot size, x-ray pulses within 2 μs with their x-ray doses in the range of several hundred rads at a meter for x-ray imaging. To achieve its performance specifications, the DARHT-II x-ray converter material is inertially confined after it turns into plasma by the heating of previous beam pulses[2]. Furthermore, the beam-target interactions, such as the focusing by backstreaming ions from desorbed gas from the target surface for the first pulse and by those from the target plasma for the subsequent pulses, and the instabilities of the beam propagating in dense plasma, should be mitigated. We will perform the ETA-II/Snowtron double pulse target experiments to validate the DARHT-II multi-pulse target concept. To simulate the DARHT-II beam-target interactions, a target plasma will be created first by striking the 1MeV, 2 kA Snowtron beam on one side of a 5 mil Ta foamed target, and then, the 6 MeV, 2 kA ETA-II beam will enter from another side to probe the target. The foam target density is 1/5 of the solid Ta density. To validate the DARHT-II target confinement concept and to ensure generating the required x-ray dose, the better characterized ETA-II beam will be used to hit the target first. We will then measure the x-ray dose created by the Snowtron beam to benchmark our hydrodynamics modeling and x-ray dose calculations.

In this paper, we discuss the Snowtron beam parameters and how much target plasma is needed to do the scaled multi-pulse target experiment in Sec. 2. In Sec. 3, the hydrodynamics modeling of the target plasma with the Snowtron beam parameters is presented. Both the 6 MeV ETA-II beam and the 1 MeV Snowtron beam will use the same final focus lens. The final focus configuration is presented in Sec. 4. The summary is in Sec. 5.

## 2 SNOWTRON PARAMETERS

Since the fourth DARHT-II pulse has to travel through the longest plasma column. It will experience the worst beam-target interaction. The double pulse experiment is designed to simulate the beam-target interactions for the third and the fourth DARHT-II pulse. The DARHT-II target plasma discussed in this section is the plasma seen by the fourth pulse (the worst case). The Snowtron injector was a predecessor of the ETA-II injector. The 1 MeV, 2 kA Snowtron injector, using a velvet cathode, will deliver a 70 ns (FWHM) long beam with a 35 ns flattop. The energy variation during the flattop is ± 1 %. The injector voltage waveform is given in Fig. 1. The Snowtron beam's normalized Lapostolle emittance is 1200 π mm-mr. To simulate the DARHT-II beam-target interactions, the ETA-II beam needs to travel through an over-densed target plasma which fully charge-neutralizes the beam and has a magnetic diffusion time of the order of 1 ns. The plasma temperature should be similar to that of the DARHT-II plasma. The temperature is linearly proportional to the energy density deposited by the beam. Assume that only the flattop portion of the Snowtron beam can deposit energy into a small spot due to chromatic aberration of the lens, and ignore the energy deposited by the rise and fall of the beam. The DARHT-II target is a 1 cm long Ta foamed target with 1/10 of the solid Ta density[2]. To achieve the temperature of the target plasma seen by the fourth DARHT-II pulse, we need to focus the Snowtron beam on the ETA-II target to a 1 mm FWHM spot.

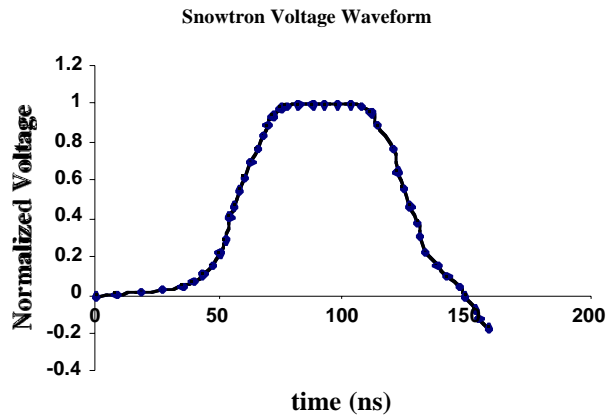

Fig.1 The Snowtron injector voltage waveform
.

The focusing effects of the target plasma and the backstreaming ions can over-focus the beam. The length of the ion channel or the plasma column needed to disrupt the final focus spot size $R$ is proportional to $R/(\gamma/I)^{1/2}$. The plasma column created by the Snowtron beam needs only 40 % of the DARHT-II plasma column length to disrupt the ETA-II beam. Since the backstreaming ion channel increases in time, its net focusing effects also increase in time. The time needed to observe the beam disruption, after the backstreaming ions are born, also follows the same scaling. For the case that the beam traveling through a pre-existing target plasma, the backstreaming ions may appear as soon as the beam arrives at the target. The beam disruption time for the ETA-II beam is only 40 % of that for the DARHT-II beam. Therefore, it is easier to observe the backstreaming ions' focusing effects on the ETA-II beam.

## 3 SNOWTRON TARGET PLASMA

We have modeled the target plasma created by the Snowtron beam using the LASNEX hydrodynamics code. We assume that the Snowtron beam has a Gaussian distribution. The energy deposited by the non-flattop of the beam is ignored. The deposited energy is distributed in the target according to a Monte Carlo calculation using the MCNP code. The third DARHT-II pulse will travel through a 1.1 cm long plasma column. According to the discussion in the previous section, the ETA-II beam has to travel through a 0.5 cm plasma column to experience similar beam-target interactions. The Snowtron plasma at the ETA-II side will have expanded approximately 0.5 cm at 1 μs after the Snowtron beam hits the target. Therefore, firing the ETA-II beam at this time can simulate the beam-target interaction for the third DARHT-II pulse. The density contours of the simulated target plasma for the third pulse are shown in Fig. 2. The foamed target is located at z = 2.3 cm in the plot. Due to the low beam energy, the large scattering by the target makes the electrons in the Snowtron beam deposit more energy at the entrance side of the target than at the exit side. Therefore, there is less plasma expansion at the exit side. The corresponding plasma density and temperature along the z-axis are shown in Figs. 3(a) and (b). The ETA-II beam's number density is $5.3 \times 10^{13}$ c.c.$^{-1}$ which is much less than that for the target plasma. Therefore, the beam is full charge neutralized by the target. The Snowtron plasma's temperature is about half of the DARHT-II plasma. The magnet diffusion times varies from 0.01 ns to 1 ns depending on the beam radius along the axis. Therefore, the interactions between the ETA-II beam and the Snowtron target plasma would be similar to that between the DARHT-II beam and and its target plasma. Similarly, to simulate the fourth DARHT-II pulse's beam-target interaction, the ETA-II beam should be fired at 1.5 μs after the Snowtron beam.

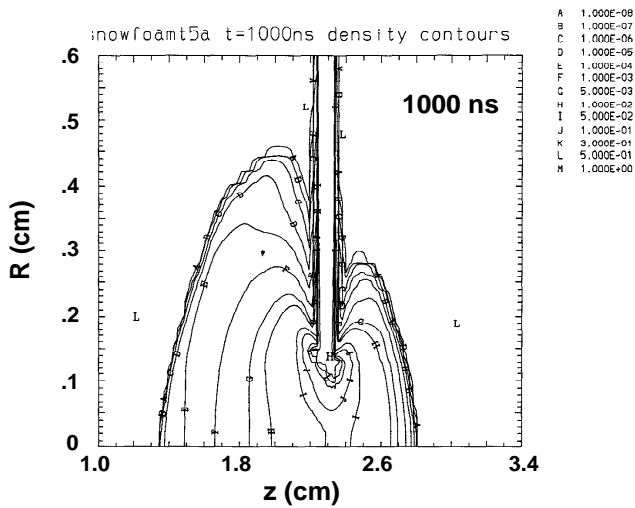

Fig. 2 The plasma density contours at 1 μs after the Snowtron beam hits the 5 mils Ta foam target with 1/5 of the solid Ta density

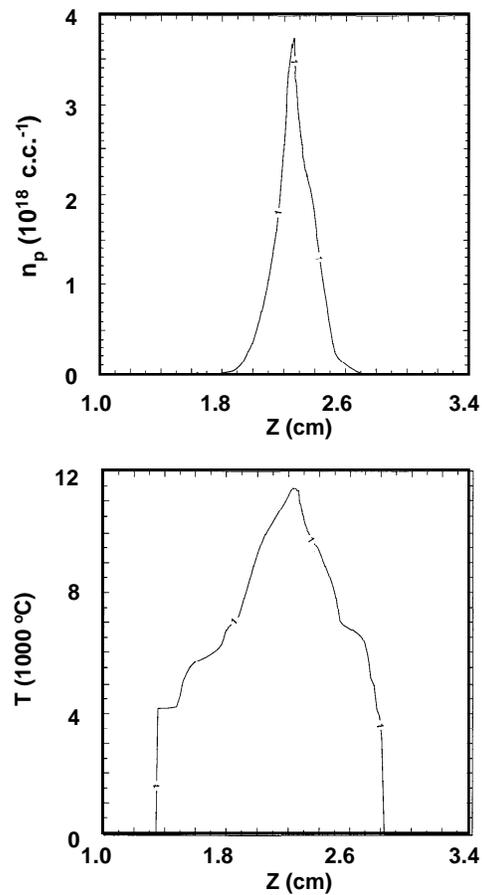

Fig. 3 The plasma (a) density and (b) temperature along the z-axis at 1 μs after the Snowtron beam hits the 5 mil Ta foam target with 1/5 of the solid Ta density.

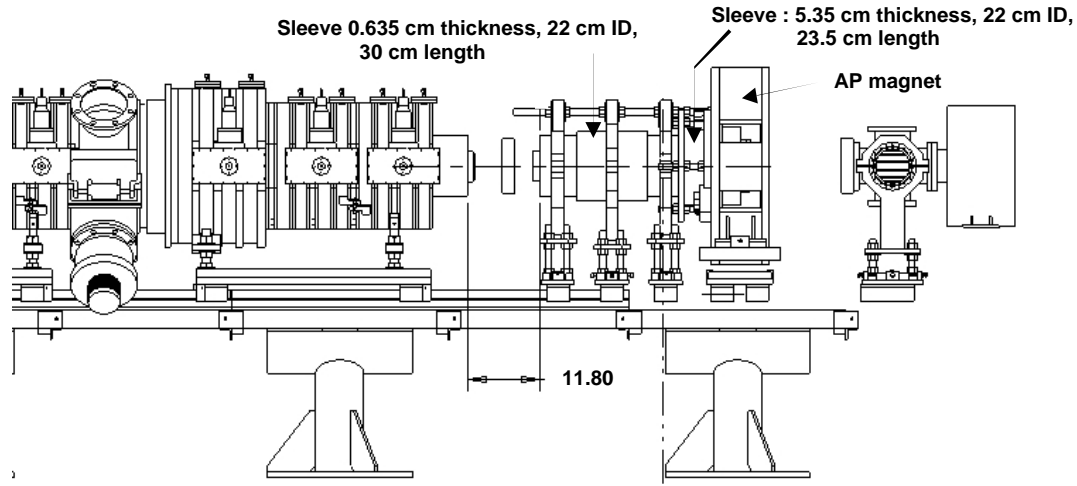

Fig. 4 The Snowtron injector and beamline configuration. The target is in the center of the AP magnet. The ETA-II beam strikes the target from the right.

## 4 FINAL FOCUS

The configuration for the Snowtron beamline is shown in Fig. 4. In order to provide enough space for diagnostics, the final focus magnet (AP magnet) for the ETA-II target experiment has a large inner bore with an ID of 39.4 cm, and the ETA-II target is located inside a focusing magnet near the center. Therefore, the same magnet will also be used to focus the Snowtron beam. Since the Snowtron beam energy is much less than the ETA-II beam energy, two iron sleeves will be inserted from the Snowtron side of the magnet to reduce the excess magnetic field. A 23.5 cm long sleeve with a thickness of 5.35 cm is located near the center of the final lens. The second sleeve, 17.5 cm away from the end of the first sleeve, is 30 cm long and 0.635 cm thick. Both sleeve's ID are 22 cm. The excitation of the final focus lens will be set by focusing the ETA-II beam to a 1 mm FWHM spot. The nominal focusing field is 4800 Gauss. A solenoid (M80) is placed between the two sleeves to match the Snowtron beam from the injector exit into the final focus region for a 1 mm FWHM spot size also. The M80 magnet is 15.4 cm in length and has an 18.2 cm ID and a 25.4 cm OD. These two sleeves and the matching lens can also be moved in and out as a unit to improve the tunability for the Snowtron beam. According the MCNP calculation, about half of the Snowtron beam electrons will be backscattered to the backward hemisphere[4]. We estimate that approximately half of these backscattered electrons would be confined by the final focus field to within the Snowtron beam radius in the final focus region. However, we do not anticipate any serious problems in focusing the Snowtron beam in the presence of these backscattered electrons. Since the sleeves and the matching solenoid are movable, we can adjust their position to allow a stronger focus field to compensate the backscattered electrons' defocusing effects.

## 5 CONCLUSIONS

We have modified the single pulse target experimental facility on the ETA-II to perform the double pulse target experiments to validate the DARHT-II multi-pulse target concept. By using the 1.15 MeV, 2 kA Snowtron beam to generate a target plasma and then firing the 6 MeV, 2 kA ETA-II beam at an appropriate time, we can simulate the beam-target interactions for various DARHT-II pulses.

## 6 ACKNOWLEDGEMENTS

We would like to thank G. Caporaso and G. Westenskow for many useful discussions.

## 7 REFERENCES

[1] S. Sampayan, et. al., this conference proceeding.
[2] M. Burns, et. al., Proc. of PAC 99, New York, N. Y. March 27 – April 2, 1999, p. 617.
[3] Y.-J. Chen, et. al., Proc. Of PAC 99, New York, NY, March 27 – April 2, 1999, p. 1827.
[4] S. Falabella, et. al., this conference proceeding.